\documentclass[12pt,preprint]{aastex}

\usepackage{amsmath,natbib,graphicx}
\usepackage{epsf}
\usepackage{fancyvrb}
\usepackage{epstopdf}
\usepackage{epsfig}
\usepackage{color}
\usepackage{lscape}
\DeclareGraphicsExtensions{.jpg,.pdf,.png,.eps,.ps}

\begin{document}
\VerbatimFootnotes

\title{Analytical investigation of the decrease in the size of the habitable zone due to limited CO$_2$ outgassing rate}
\shorttitle{Carbon Cycle}

%%%%%%%%%%%%%%%%%%%%%%%%%%%
% AUTHORS AND AFFILIATIONS
%%%%%%%%%%%%%%%%%%%%%%%%%%%

\author{Dorian~S.~Abbot\altaffilmark{1}}
\altaffiltext{1}{Department of the Geophysical Sciences, University of
  Chicago, 5734 South Ellis Avenue, Chicago, IL 60637}
\shortauthors{Abbot}

\email{abbot@uchicago.edu}

%%%%%%%%%%%%%%%%%%%%%%%%%%%
% ABSTRACT
%%%%%%%%%%%%%%%%%%%%%%%%%%%

\begin{abstract}
  The habitable zone concept is important because it focuses the
  scientific search for extraterrestrial life and aids the planning of
  future telescopes. Recent work has shown that planets near the outer
  edge of the habitable zone might not actually be able to stay warm
  and habitable if CO$_2$ outgassing rates are not large enough to
  maintain high CO$_2$ partial pressures against removal by silicate
  weathering. In this paper I use simple equations for the climate and
  CO$_2$ budget of a planet in the habitable zone that can capture the
  qualitative behavior of the system. With these equations I derive an
  analytical formula for an effective outer edge of the habitable
  zone, including limitations imposed by the CO$_2$ outgassing rate. I
  then show that climate cycles between a Snowball state and a warm
  climate are only possible beyond this limit if the weathering rate
  in the Snowball climate is smaller than the CO$_2$ outgassing rate
  (otherwise stable Snowball states result). I derive an analytical
  solution for the climate cycles including a formula for their period
  in this limit. This work allows us to explore the qualitative
  effects of weathering processes on the effective outer edge of the
  habitable zone, which is important because weathering
  parameterizations are uncertain.
\end{abstract}
 \keywords{planets and satellites: atmospheres, astrobiology}

\bigskip\bigskip

\section{Introduction}

The habitable zone is defined as the region around a star where a
planet with CO$_2$ and H$_2$O as its main greenhouse gases can
support liquid water at its surface \citep{Kasting93}. The habitable
zone is relatively wide because of the silicate-weathering feedback
\citep{Walker-Hays-Kasting-1981:negative}. Silicate-weathering is a
geological process that removes CO$_2$ from the atmosphere and a
negative (stabilizing) feedback is possible because this process is
believed to run faster at higher temperatures and higher CO$_2$
partial pressures \citep{Pierrehumbert:2010-book}. The inner edge of
the habitable zone is set by the moist or runaway greenhouse
\citep{Kasting88,Nakajima92,Goldblatt:2012}, which should not be
influenced by the details of the silicate-weathering feedback since
the CO$_2$ should have been drawn down to low levels when they
occur. On the other hand, calculations of the outer edge of the
habitable zone generally assume that the silicate-weathering feedback
can maintain CO$_2$ at arbitrarily high levels. The outer edge is then
marked by some threshold where adding CO$_2$ to the atmosphere no
longer provides additional warming, for example if additional CO$_2$
increases Rayleigh scattering more than it increases greenhouse
warming \citep{Kasting93,Kopparapu:2013}. This picture essentially
considers the asymptotic limit of unlimited CO$_2$ outgassing
capacity.

More recently it has been recognized that for finite CO$_2$ outgassing
rates the maximum CO$_2$ that a planet can achieve may be lower than
the CO$_2$ needed to keep that planet habitable at the outer edge of
the habitable zone \citep{Tajika2007,Kadoya:2014kd}. This leads to
what I will call the ``effective outer edge of the habitable zone,''
which will depend on the CO$_2$ outgassing rate and the functioning of
silicate weathering on the planet. Beyond the effective outer edge of
the habitable zone, a planet may experience cycles
\citep{Menou2015,haqq2016limit} between a globally frozen, Snowball
Earth climate \citep{Kirschvink92,Hoffman98} and a habitable climate.
Understanding and constraining the effective outer edge of the
habitable zone is critical because it determines our estimate of the
fraction of stars that host an Earth-like planet
\citep{Petigura:2013,Kopparapu:2013fp}, which is essential for
planning future telescopes that would observe such planets.

The purpose of this paper is to investigate what determines the
position of the effective outer edge of the habitable zone and what
happens beyond the effective outer edge. The behavior of a planet in
this regime is largely determined by its uncertain silicate-weathering
behavior. Instead of viewing this an an obstacle, I will exploit it to
make progress on the problem. The importance and uncertainty of
weathering justifies the use of a simple climate model since the
errors associated with making the grave assumptions that such a model
entails pale in comparison to the uncertainty in weathering. It also
allows me to use a relatively simple weathering parameterization that
captures the expected qualitative behavior. I will use the simplifying
assumptions to derive useful formulae. This will allow us to
understand issues such as which processes determine whether the
effective outer edge of the habitable zone is inside the traditional
outer edge of the habitable zone and how much the parameters
associated with these processes would need to be changed from our best
estimates of their values to change the qualitative behavior of the
system. Moreover, I am able to determine the conditions for climate
cycles to occur beyond the effective outer edge of the habitable zone,
an analytical solution for these cycles, and a formula for the period
of the cycles. Using more complicated climate and weathering models
will alter quantitative results, but is unlikely to alter the
qualitative dependencies on parameters that my formulae give. This
work therefore complements recent work by \citet{Menou2015} and
\citet{haqq2016limit}, who used more complicated radiative and climate
models, but only explored a few values of weathering parameters.

I will neglect sophisticated radiative
\citep{Kopparapu:2013fp,Goldblatt:2013} and 3D calculations
\citep{Leconte:2013gv,leconte2013increased,yang2013,yang2014,Wolf:2014,wolf2015evolution}
and consider the following linearized, zero-dimensional model of
planetary climate, similar to that used by \citep{abbot12-weathering}:
\begin{equation}
  C\frac{dT}{dt} = \frac{S}{4}(1-\alpha(T))-\frac{S_0}{4}(1-\alpha_o)-a(T-T_0)+b \log\left(\frac{P}{P_0}\right),
\label{eq:climate_model}
\end{equation}
where $a$ and $b$ are constants, $T$ is the global mean temperature,
$T_0$ is the temperature of the reference state, $P$ is the
atmospheric partial pressure of CO$_2$, $P_0$ is the atmospheric
partial pressure of CO$_2$ in the reference state, $S$ is the stellar
flux, $S_0$ is the stellar flux in the reference state, $C$ is the
heat capacity in units of J m$^{-2}$~$^\circ$C$^{-1}$, $\alpha(T)$ is
the temperature-dependent planetary albedo (reflectivity), and
$\alpha_0$ is the albedo of the reference state.
Table~\ref{tab:params} contains a list model parameters and their
standard values. The values I use here are drawn from
\citet{abbot12-weathering}, \citet{Menou2015}, and
\citet{haqq2016limit}, but it is important to emphasize that the
arguments in this paper do not depend on the exact values chosen.

I will let the albedo be specified by
\begin{equation}
\alpha(T) = \begin{cases}
  \alpha_w,\   T\geq T_i,\\
  \alpha_c,\  T<T_i,
            \end{cases}
\label{eq:albedo}
\end{equation}
where $\alpha_w$ is the albedo of the warm climate state; $\alpha_c$
is the albedo of the cold and icy, ``Snowball'' climate state; and
$T_i$ is the temperature at which the planet transitions between the
two climate states. Equation~(\ref{eq:albedo}) assumes a step-wise
transition in albedo, which neglects the effects of spatial resolution
\citep{yang2011a,yang2011b,yang2012,voigt11,voigt12-dynamics}.
Nevertheless, it allows us to make easier analytical progress and does
not alter the qualitative behavior of the system unless multiple
Snowball-like climate states possible
\citep{Abbot-et-al-2011:Jormungand,rose2015stable}, which is a
possibility we will not consider here. As we will see below,
introducing a smoothed albedo transition only introduces a repulsing
fixed point in some situations that the climate could not exist stably
in. Finally, if we assume that the reference climate state is warm,
then $\alpha_0=\alpha_w$.

The partial pressure of CO$_2$
is determined by outgassing and weathering
\begin{equation}
  \frac{dP}{dt} = V - W_0 e^{k(T-T_0)}\left( \frac{P}{P_0} \right)^\beta,
\label{eq:co2_bal}
\end{equation}
where $V$ is the CO$_2$ outgassing rate, $W_0$ is the rate of removal
of CO$_2$ from the atmosphere by silicate weathering in the reference
climate state, $k$ is a rate constant for the increase in weathering
with temperature, and $\beta$ is an exponent that determines how
strongly weathering depends on atmospheric CO$_2$ partial pressure.
\citet{west2005tectonic} used an analysis of river catchments to
estimate that $k=0.11 \pm 0.04$~$^\circ$C$^{-1}$ (1-$\sigma$ error).
Based on that study, the plausible range for $k$ is roughly 0--0.2. I
will use a standard value of $k$=0.1 and vary $k$ over this range.
$\beta$ could be zero if land plants concentrate CO$_2$ in the soil at
the same level regardless of the atmospheric CO$_2$ concentration
\citep{Pierrehumbert:2010-book}, and theoretical arguments suggest
$\beta$ has a maximum of 1 \citep{Berner:1994p3295}. I will use a
standard value of $\beta$=0.5
\citep{Berner:1994p3295,Pierrehumbert:2010-book,abbot12-weathering,Menou2015,haqq2016limit},
but consider variations within the plausible range. Depressurization
caused by glacial unloading can cause temporary increases in the
CO$_2$ outgassing rate \citep{huybers2009feedback}. I neglect this
affect and take the CO$_2$ outgassing rate to be independent climate
state here, which is appropriate for longterm average behavior. Note
also that in section~\ref{sec:cycles} we will consider $W_0$=0 when
$T<T_i$.

For the purposes of this paper I am assuming that weathering follows a
similar parameterization on land and at the seafloor, but it should be
understood that the weathering behavior of an ocean planet would
likely be quite different from that of a planet with an Earth-like
land fraction \citep{abbot12-weathering}. The weathering
parameterization in Equation~(\ref{eq:co2_bal}) is similar to that
used by other authors
\citep{Berner:1994p3295,Berner2004,Pierrehumbert:2010-book,abbot12-weathering,Menou2015,haqq2016limit},
but I have dropped the relatively weak dependence on temperature that
is often included to represent changes in precipitation with
temperature. The qualitative behavior of weathering as a function of
temperature is captured by Equation~(\ref{eq:co2_bal}) without this
additional complication. Given the ad hoc nature of all weathering
parameterizations, it is reasonable to choose the simplest
parameterization that gives the expected qualitative behavior of
weathering processes given the objectives of this paper.

Equations~(\ref{eq:climate_model})-(\ref{eq:co2_bal}) define the
system. The plan for analyzing them is as follows. First we will
consider the warm (habitable) state (section \ref{sec:warm}). I will
set the albedo (Equation~(\ref{eq:albedo})) to its warm state value,
set the time derivatives in Equations~(\ref{eq:climate_model}) and
(\ref{eq:co2_bal}) to zero, and solve the system for conditions when
the temperature is high enough for the warm state to exist. This will
allow us to put bounds on the existence of the warm state, that is,
define an effective outer edge of the habitable zone that may be more
restrictive than the traditional outer edge. Next I will find
nullclines of the system defined by Equations~(\ref{eq:climate_model})
and (\ref{eq:co2_bal}) (that is, lines where the time derivatives
equal zero), find their intersections (fixed points, or solutions of
the system), and determine the stability of these fixed points.
Physically, this will reveal that if Equation~(\ref{eq:co2_bal}) is
followed as is, enough weathering occurs that the system settles into
a stable Snowball state when the warm climate state ceases to exist,
rather than into climate cycles between Snowball and warm conditions.
I will then show that climate cycles do occur if we set weathering to
zero in the Snowball state in Equation~(\ref{eq:co2_bal})
(section~\ref{sec:cycles}). I will find analytical solutions for the
components of these cycles and derive an analytical formula for their
period that can predict well results from the more intricate model of
\citep{Menou2015}. I will then discuss these results in
section~\ref{sec:discussion} and conclude in
section~\ref{sec:conclusions}.

\begin{table}[h!]
  \caption{A list of the model parameters, their descriptions, and the standard values I use for them. The parameter values in this table are taken from
    \citet{abbot12-weathering}, \citet{Menou2015}, and \citet{haqq2016limit}. Note that, following \citet{haqq2016limit}, I use a value of $W_0$ ten times higher than \citet{Menou2015}. I perform sensitivity analyses where I vary $k$ and $\beta$.}
\label{tab:params}
\centering
\begin{tabular}{lll}
  Parameter & Description & Standard Value \\
  \hline
  $S_0$ & reference state stellar flux & 1365.0~W~m$^{-2}$ \\
  $S$ & stellar flux & variable \\
  $T_0$ & reference state temperature &  15.0$^\circ$C \\
  $P_0$ & reference state partial pressure CO$_2$ &  3$\times10^{-4}$  bars \\
  $a$ & slope of planetary infrared emission with temperature & 2.0~W~m$^{-2}$~$^\circ$C$^{-1}$ \\
  $b$ & slope of planetary infrared emission with logarithm of CO$_2$ & 10.0~W~m$^{-2}$ \\
  $\alpha_0$ & reference state albedo & 0.3 \\
  $\alpha_w$ & warm state albedo & 0.3 \\
  $\alpha_c$ & cold state albedo & 0.6 \\
  $T_i$ & albedo transition temperature & -10$^\circ$C \\
  $C$ & planetary heat capacity  & 2$\times10^8$~J~m$^{-2}$~$^\circ$C$^{-1}$ \\
  $V$ & CO$_2$ outgassing rate & variable \\
  $W_0$ & reference state CO$_2$ weathering rate & 70~bars~Gyr$^{-1}$ \\
  $k$ &  weathering-temperature rate constant & 0.1~$^\circ$C$^{-1}$ \\
  $\beta$ & weathering-CO$_2$ power law exponent & 0.5 
\end{tabular}
\end{table}

\section{Conditions on the existence of a habitable climate state}
\label{sec:warm}

In this section we will consider the conditions that allow the carbon
cycle to maintain a planet in the warm climate state, which is
necessary for the planet to be considered habitable in the traditional
sense. This requires a steady-state solution, so we can set the time
derivatives in Equations~(\ref{eq:climate_model}) and
(\ref{eq:co2_bal}) to zero. We can solve this system to find
\begin{equation}
  T_w -T_0= \frac{b \log \left( \frac{V}{W_0} \right) +\frac{\beta}{4}(S-S_0)(1-\alpha_w)}{kb+a\beta},
\label{eq:T_w}
\end{equation}
where $T_w$ is the warm state temperature. We could rewrite
Equation~(\ref{eq:T_w}) in non-dimensional form, but I will leave it,
and subsequent equations, in dimensionful form to make them more
accessible physically. From Equation~(\ref{eq:T_w}) we can see that
increasing the CO$_2$ outgassing rate logarithmically increases the
warm state temperature, and that increased outgassing warms the warm
state more if the climate is more sensitive to CO$_2$ (higher $b$),
provided $\beta \neq 0$. We can also see that the influence of stellar
flux on the warm state temperature depends on $\beta$, the
weathering-CO$_2$ power-law exponent. If $\beta=0$, then changing the
stellar flux has no effect on the warm state temperature as long as
the stellar flux is high enough that $T_w \geq T_i$, so that the warm
state can exist. Instead, the warm state temperature is determined by
the outgassing rate. As $\beta$ increases, the warm climate state
becomes increasingly sensitive to changes in stellar flux.

We can solve for the warm state CO$_2$ partial pressure ($P_w$) as follows
\begin{equation}
  \log \left(\frac{P_w}{P_0}\right) = \frac{ a \log \left( \frac{V}{W_0} \right) + \frac{k}{4}(S_0-S)(1-\alpha_w)}{kb+a\beta}.
\label{eq:P_w}
\end{equation}
The warm state CO$_2$ partial pressure has a power law dependence on
the CO$_2$ outgassing rate. The exponent depends most strongly on $a$,
which determines the increase in outgoing longwave radiation with
temperature. The more that increasing the warm state temperature by a
given amount increases longwave cooling, the higher the warm state
CO$_2$ must be to maintain that temperature at a given CO$_2$
outgassing rate. As expected, the warm state CO$_2$ partial pressure
is exponentially lower if the stellar flux is higher. This effect is
 mediated mainly by $k$, the parameter that determines the
increase in weathering with temperature. If $k$ is higher, then a given
warming from an increase in stellar flux causes the weathering to
increase more, and draws down the CO$_2$ more.

Using Equation~(\ref{eq:T_w}), we can solve for the coldest possible
warm state by setting $T_w=T_i$. This is a very important condition
because habitability would be lost if anything were to cool the
climate when it is in the coldest possible warm state. We can solve
for the stellar flux at which the coldest possible warm state exists
($S^\ast$), which we can think as the effective outer edge of the
habitable zone, as follows
\begin{equation}
  S^\ast = S_0-\left(\frac{4}{1-\alpha_w}\right)\left(\frac{b}{\beta}\log\left(\frac{V}{W_0}\right)+k(T_0-T_i)\left(\frac{b}{\beta}+\frac{a}{k}\right)\right).
\label{eq:s_star}
\end{equation}
A lower value of $S^\ast$ means that the stellar flux must be
decreased more to reach the coldest possible warm state, which means
that the warm climate state exists in more of the habitable zone. The
ideal situation for habitability is when $S^\ast$ is decreased enough
that it is smaller than the outer edge of the habitable zone, the
point at which some other process, such as Rayleigh scattering,
prevents CO$_2$ from warming the planetary surface. If this is the
case then the silicate-weathering feedback can keep a planet habitable
throughout the entire habitable zone.

\begin{figure}[h!]
\begin{center}
\epsfig{file=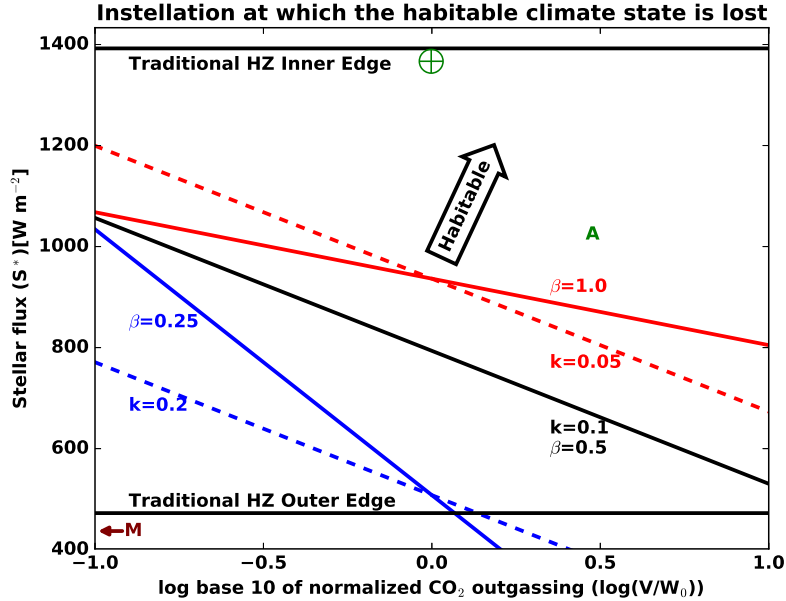, width=0.7\textwidth} 
\end{center}
\caption{The stellar flux at which the habitable climate state is lost
  (the effective outer edge of the habitable zone) as a function of
  the logarithm of the normalized CO$_2$ outgassing (black line for
  standard parameters). The different curves show the behavior for
  different values of the weathering-CO$_2$ power law exponent
  ($\beta$, solid red and blue lines) and the weathering-temperature
  rate constant ($k$, dashed red and blue lines), both of which are
  relatively unconstrained. Red lines indicate changes to parameters
  that restrict the region where the habitable climate state can exist
  and blue lines indicate changes to parameters that expand this
  region. The traditional inner and outer edges of the habitable zone
  from \citet{Kopparapu:2013} are at the top and bottom of this plot.
  The symbol $\bigoplus$ represents modern Earth, the ``A'' represents
  Earth 3.8 Gyr ago assuming 75\% modern Earth's stellar flux
  \citep[i.e., insolation,][]{GOUGH:1981p2371} and a CO$_2$ outgassing
  rate of three times modern, and the ``M'' represents Mars 3.8 Gyr
  ago assuming a CO$_2$ outgassing rate of 1 bar Gyr$^{-1}$
  \citep{grott2011volcanic}, which would plot to the left of the
  minimum CO$_2$ outgassing rate shown here. The CO$_2$ outgassing
  rate on early Earth is not well-constrained, but probably was higher
  than modern \citep{dasgupta2013ingassing}.}
\label{fig:outer_outgassing2}
\end{figure}

The first thing we can note from Equation~(\ref{eq:s_star}) is that if
the albedo of the warm state is smaller, the effective outer edge of
the habitable zone will be further out. Since absorption of stellar
flux by atmospheric water vapor should be larger for planets orbiting
smaller, redder stars \citep{Kasting93}, the effective outer edge of
the habitable zone is less likely to restrict the traditional
habitable zone for M-stars than for G-stars and more likely for
F-stars, which is consistent with what has been recently found by
\citet{haqq2016limit}. Equation~(\ref{eq:s_star}) also tells us that
the effective outer edge of the habitable zone depends only
logarithmically on the CO$_2$ outgassing rate. This is important
because it tells us that large changes in the CO$_2$ outgassing rate
are necessary to significantly change the effective outer edge. For
example, increasing the CO$_2$ outgassing rate by a factor of ten only
decreases the effective outer edge from 794~W~m$^{-2}$ to
530~W~m$^{-2}$, for our standard parameters
(Figure~\ref{fig:outer_outgassing2}). Changing the weathering
parameters can have a much larger effect, which we can investigate by
doing a sensitivity analysis in which we vary them. For example,
either increasing the weathering-temperature rate constant ($k$) by a
factor of two or decreasing the weathering-CO$_2$ power law exponent
($\beta$) by a factor of two causes a similar reduction in the
effective outer edge as increasing the CO$_2$ outgassing by a factor
of ten (Figure~\ref{fig:outer_outgassing2}).

Changing $k$ and changing $\beta$ have qualitatively different effects
on the response of $S^\ast$ to changes in the CO$_2$ outgassing rate
(Figure~\ref{fig:outer_outgassing2}). Decreasing $\beta$ increases the
slope of $S^\ast$ as a function of $\log\left(\frac{V}{W_0}\right)$,
which means that increasing the CO$_2$ outgassing rate by a given
amount extends the effective habitable zone further
(Equation~(\ref{eq:s_star})). This is because the temperature becomes
more sensitive to the CO$_2$ outgassing rate when $\beta$ is smaller
(Equation~(\ref{eq:T_w})). Alternatively, changing $k$ changes the
offset of the $S^\ast$ versus $\log\left(\frac{V}{W_0}\right)$ line,
but not the slope. A higher value of $k$ means that the temperature
has to change less to cause the same change in weathering rate. This
allows the stellar flux to be dropped to a lower value before the warm
state temperature reaches the temperature at which the warm state is
lost ($T_w=T_i$). 

For reference I have plotted the estimated positions
of modern Earth, Earth 3.8~Gyr ago, and Mars 3.8~Gyr on
Figure~\ref{fig:outer_outgassing2}. The simple model used here
indicates that early and modern Earth are safely inside the effective
outer edge of the habitable zone, whereas early Mars would have been
beyond the effective outer edge of the habitable zone even if it were
within the traditional habitable zone. This is consistent with the
histories of the two planets if we assume that fluvial features on
early Mars were episodic
\citep{Wordsworth:2013fk,halevy2014episodic,kite2015stratigraphy}.
That said, it should be understood that specific conclusions like
these depend on details of weathering parameterizations, and the
purpose of this paper is to expose the qualitative effects of
weathering, rather than to try to answer detailed questions.

\begin{figure}[h!]
\begin{center}
\epsfig{file=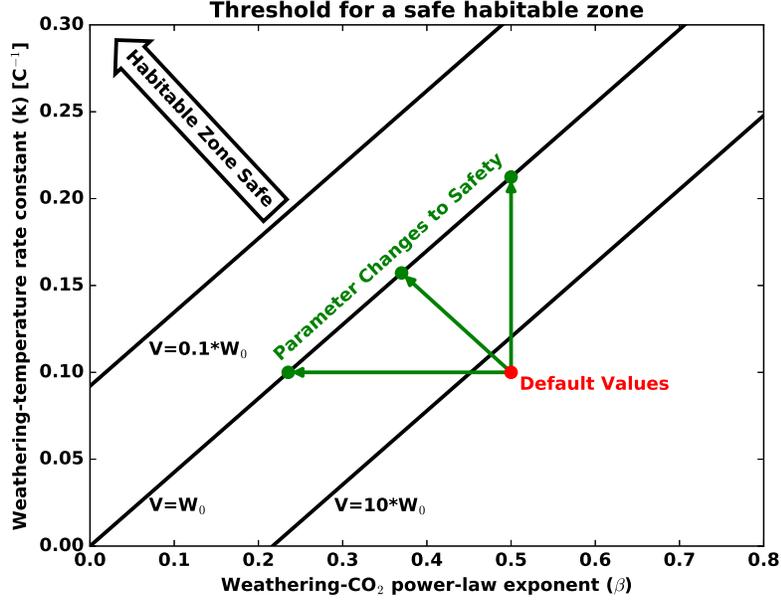, width=0.7\textwidth} 
\end{center}
\caption{Lines of the weathering-temperature rate constant ($k$) as a
  function of the weathering-CO$_2$ power law exponent ($\beta$) at
  which the effective outer edge of the habitable zone equals the
  traditional outer edge. These lines are plotted for three different
  values of the CO$_2$ outgassing rate (0.1, 1, and 10 times modern
  Earth's value). If the $k$ and $\beta$ combination is to the upper
  left of this plot relative to a given line, then CO$_2$ outgassing
  does not limit habitability and the full habitable zone is saved.}
\label{fig:outer_outgassing}
\end{figure}

Another way we can think about this is to consider the set of
weathering parameters that would make the effective outer edge of the
habitable zone correspond to the traditional outer edge of the
habitable zone. If we denote the traditional outer edge of the
habitable zone as $S^\ast_{out}$, we can rewrite
Equation~(\ref{eq:s_star}) as
\begin{equation}
  k = - \frac{\log\left(\frac{V}{W_0}\right)}{T_0-T_i}+\frac{\beta}{b}\left(\frac{\frac{1}{4}(S_0-S^\ast_{out})(1-\alpha_w)}{T_0-T_i}-a\right).
\label{eq:outer_edge}
\end{equation}
Equation~(\ref{eq:outer_edge}) represents a series of lines of $k$ as
a function of $\beta$ for different values of the CO$_2$ outgassing
rate. If $k$ is larger or $\beta$ is smaller than the line for a
particular CO$_2$ outgassing rate, then the effective habitable zone
is just as large as the traditional habitable zone for this CO$_2$
outgassing rate. Figure~\ref{fig:outer_outgassing} shows three such
curves for different CO$_2$ outgassing rates. For the default values
of $k$ and $\beta$, some of the habitable zone would be lost even if
the CO$_2$ outgassing rate were ten times higher than modern Earth's
($\frac{V^\ast}{W_0}=16.6$ would be required for the effective outer
edge to equal the traditional outer edge). However, relatively small
changes to $k$, $\beta$, or some combination of the two can save the
habitable zone even for modern Earth's CO$_2$ outgassing rate
(Figure~\ref{fig:outer_outgassing}). The parameters $k$ and $\beta$
are uncertain and may vary between planets, but the expressions in
this section could be used for a probabilistic estimate of planetary
habitability if appropriate priors are used for $k$ and $\beta$.

\section{What happens when the habitable climate state ceases to exist}
\label{sec:stable_snow}

If the stellar flux or CO$_2$ outgassing rate is lowered enough that
the warm climate state no longer exists, the planet enters a Snowball
state and the albedo increases according to
Equation~(\ref{eq:albedo}). It is an open question how weathering
would behave in a Snowball state. \citet{Menou2015} assumed that
weathering would go to zero due to lack of rain (liquid water), and we will
consider this case in section~\ref{sec:cycles}. The possibility
remains, however, that weathering could continue to occur under
wet-based ice sheets or at the seafloor \citep{Lehir08}. For
illustrative purposes, we will continue to use
Equation~(\ref{eq:co2_bal}) to characterize weathering in the Snowball
state, but it should be understood that either subglacial or seafloor
weathering could lead to different parameterizations. As we will see,
however, the important point is that if some CO$_2$-dependent
weathering can occur in the Snowball state, and it causes weathering
to increase enough to balance CO$_2$ outgassing before the Snowball
deglaciates, then it is possible for this state to be a stable
solution of the system.

We can solve for the temperature tendency nullcline (the line on which
$\frac{dT}{dt}$=0 in Equation~(\ref{eq:climate_model})), which results
in
\begin{equation}
  \log\left(\frac{P}{P_0}\right) = \begin{cases}
    \frac{1}{b}\left(a(T-T_0)+\frac{1}{4}S_0(1-\alpha_w)-\frac{1}{4}S(1-\alpha_w)\right),\ T\geq T_i\\
    \frac{1}{b}\left(a(T-T_0)+\frac{1}{4}S_0(1-\alpha_w)-\frac{1}{4}S(1-\alpha_c)\right),\ T<T_i.
\end{cases}
\label{eq:T_nullcline}
\end{equation}
Equation~(\ref{eq:T_nullcline}) describes two lines of
$\log\left(\frac{P}{P_0}\right)$ as a function of $T-T_0$, each with a
slope of $\frac{a}{b}$. The colder solution has a larger vertical
offset. This is because more CO$_2$ would be required to keep the cold
state at a given temperature than the warm state because the albedo is
higher in the cold state (although they can never actually exist at
the same temperature). Similarly, we solve for the CO$_2$ partial
pressure tendency nullcline (the line on which $\frac{dP}{dt}$=0 in
Equation~(\ref{eq:co2_bal})) to find
\begin{equation}
  \log\left(\frac{P}{P_0}\right) = \frac{1}{\beta}\log\left(\frac{V}{W_0}\right)-\frac{k}{\beta}(T-T_0).
\label{eq:P_nullcline}
\end{equation}
Equation~(\ref{eq:P_nullcline}) describes a line of
$\log\left(\frac{P}{P_0}\right)$ as a function of $T-T_0$ with a slope
of $-\frac{k}{\beta}$. Since the slope is negative, there will always
be at least one intersection of the two nullclines, which will be a
steady-state of the system. As the CO$_2$ outgassing rate ($V$)
increases, the intercept of Equation~(\ref{eq:P_nullcline}) is
increased and the solution becomes warmer. For high values of $V$ only
the warm state is a steady state, and for low values only the cold
state is a steady state. For intermediate values of $V$ it is possible
to have both states to be a steady state of the system.

We already solved for the warm state temperature and CO$_2$ in
Equations~(\ref{eq:T_w}) and (\ref{eq:P_w}). We can now solve for the
cold state temperature ($T_c$) and CO$_2$ ($P_c$) using the cold
branch of Equation~(\ref{eq:T_nullcline})
\begin{align}
\label{eq:T_c}
  T_c-T_0=\frac{b \log \left( \frac{V}{W_0} \right) +\frac{\beta}{4}S(1-\alpha_c)-\frac{\beta}{4}S_0(1-\alpha_w)}{kb+a\beta}, \\
 \log \left(\frac{P_c}{P_0}\right) = \frac{ a \log \left( \frac{V}{W_0} \right) + \frac{k}{4}S_0(1-\alpha_w)-\frac{k}{4}S(1-\alpha_c)}{kb+a\beta}.
\label{eq:P_c}
\end{align}

The final point of interest is to determine the stability of the
fixed points described by Equations~(\ref{eq:T_w}), (\ref{eq:P_w}),
(\ref{eq:T_c}), and (\ref{eq:P_c}). We can do this by evaluating the
Jacobian ($J$) of the system at the fixed points
\begin{equation}
J = \begin{bmatrix}
\frac{\partial}{\partial T} \frac{dT}{dt} & \frac{\partial}{\partial P} \frac{dT}{dt} \\
\frac{\partial}{\partial T} \frac{dP}{dt} & \frac{\partial}{\partial P} \frac{dP}{dt} 
\end{bmatrix}
=
\begin{bmatrix}
-\frac{a}{C} & \frac{b}{CP} \\
-k W_0 e^{k(T-T_0)}\left(\frac{P}{P_0}\right)^\beta & - \frac{W_0}{P_0}\beta e^{k(T-T_0)}\left(\frac{P}{P_0}\right)^{\beta-1}
\end{bmatrix}.
\label{eq:jacobian}
\end{equation}
For both the warm and cold states, the trace of the Jacobian
($\tau$) is less than zero and its determinant ($\Delta$) is greater
than zero. This means that the warm and cold states are always
attracting \citep{Strogatz-1994:nonlinear}. $\tau^2-4\Delta$ is
generally positive, which means that the fixed points will usually be
stable nodes, although it is possible for them to be stable spirals
for some parameter combinations.

\begin{figure}[h!]
\begin{center}
\epsfig{file=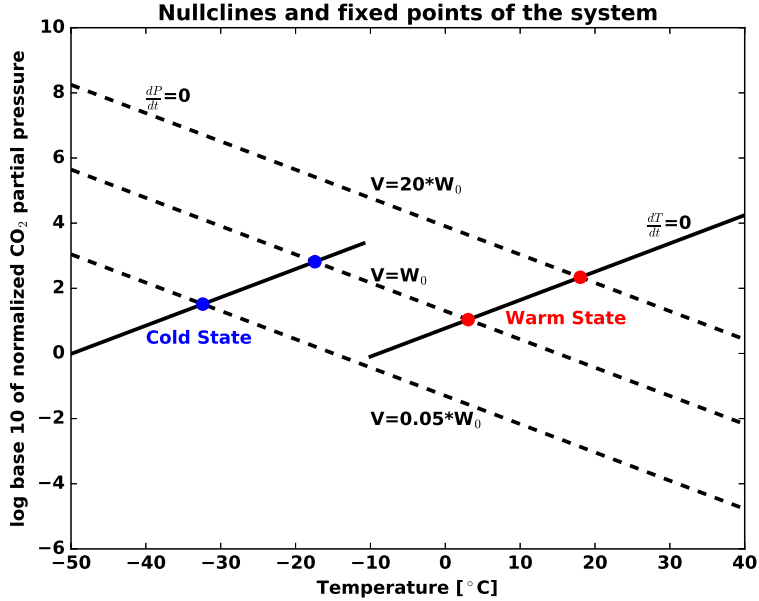, width=0.7\textwidth} 
\end{center}
\caption{This plot shows nullclines of the system, where either the temperature tendency is zero ($\frac{dT}{dt}$=0, solid lines) or the CO$_2$ partial pressure tendency is zero ($\frac{dP}{dt}$=0, dashed lines). Three pressure partial pressure tendency nullclines are shown: for a CO$_2$ outgassing rate of 0.05, 1, and 20 times modern Earth's. The intersections of these nullclines represent fixed points of the system, which are all attractors. Warm climate state fixed points are plotted in red and cold climate state fixed points are plotted in blue. For higher CO$_2$ outgassing rates only the warm state exists, for lower rates only the cold state exists, and at intermediate rates both the warm and cold climate states exist. The stellar flux is 80\% of modern Earth's in this figure.}
\label{fig:nullclines}
\end{figure}

I have plotted what we have learned about the nullclines and their
intersections for a representative set of parameters in
Figure~\ref{fig:nullclines}. This plot shows how intersections of the
nullclines lead to steady states, and how at least one intersection
will always occur since the CO$_2$ partial pressure tendency nullcline
has a negative slope and the two temperature tendency nullclines have
a positive slope. Note that although the simplicity of the model we
are considering constrains the nullclines to be linear, we would
expect similar, but potentially nonlinear, behavior from a more
complicated model. For example, if we used a more sophisticated
radiative transfer model for the climate calculations
\citep[e.g.,][]{Kopparapu:2013} it would lead to curvature in the
temperature tendency nullclines, but no change in the topology of the
system.

Because we have assumed a discontinous albedo transition
(Equation~(\ref{eq:albedo})) stable fixed points appear and disappear
 in isolation in Figure~\ref{fig:nullclines}. If we had instead assumed
a smoothed albedo transition, for example smoothed with a hyperbolic
tangent function, the two temperature tendency nullclines would
smoothly join together. In this case, there would always be an
unstable saddle fixed point between the two attracting fixed points
representing the warm and Snowball climate states when both attracting
states exist at the same CO$_2$ outgassing rate. If the CO$_2$
outgassing rate were changed sufficiently, the saddle fixed point
would merge with either of the attracting fixed points in a saddle
node bifurcation, rather than the attracting fixed point just
disappearing as it does in the discontinous albedo system.

\begin{figure}[h!]
\begin{center}
\epsfig{file=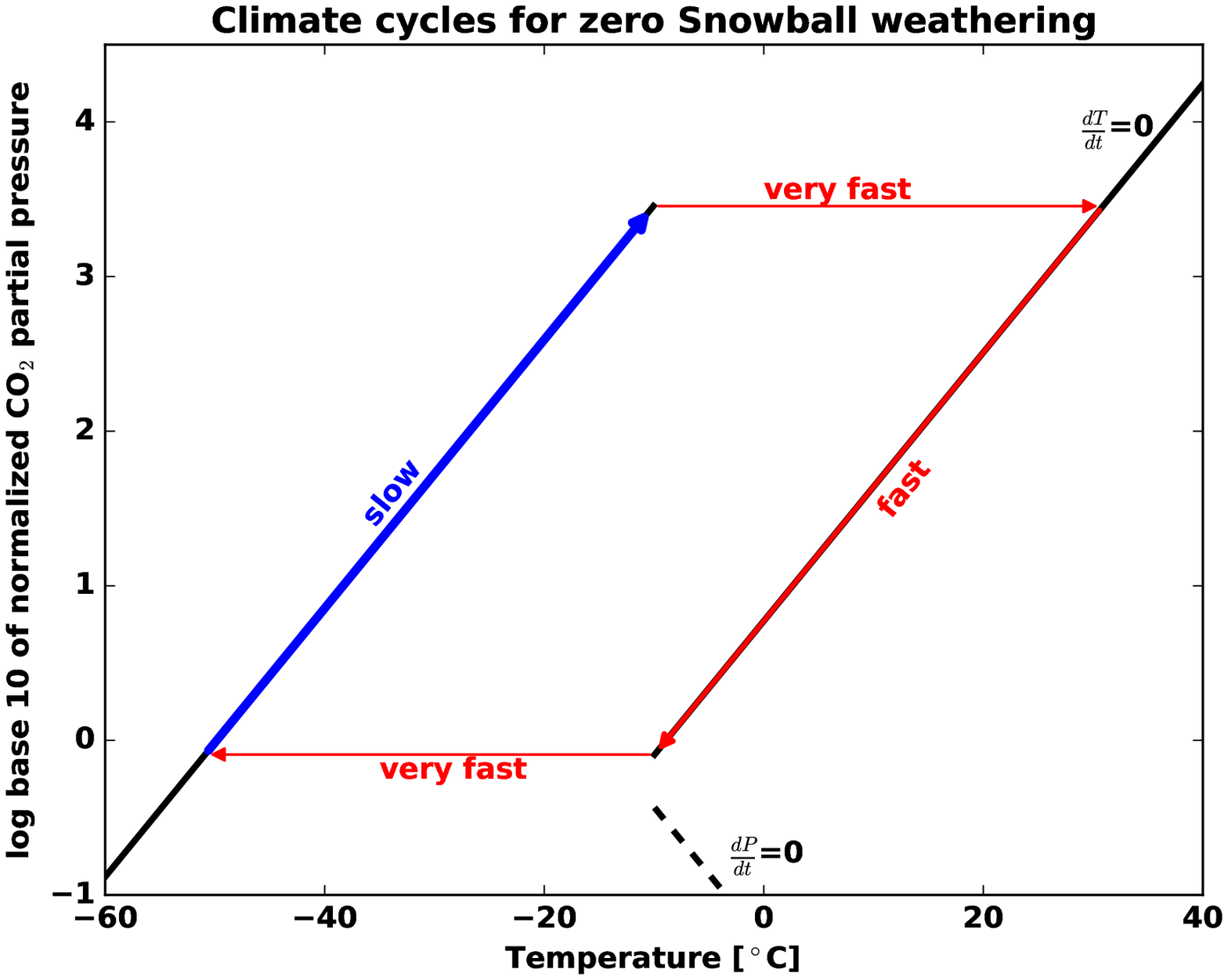, width=0.7\textwidth} 
\end{center}
\caption{Plot of the climate cycles that occur when both (1) the CO$_2$ outgassing rate is too low to achieve a warm climate steady state and (2) the weathering rate is set to zero when the temperature is less than the Snowball transition temperature ($T<T_i$). This plot is similar to Figure~\ref{fig:nullclines}, but the CO$_2$ partial pressure nullcline ends for $T<T_i$ because there is no way for the time derivative of the CO$_2$ partial pressure to be zero if the weathering rate is zero. There is no steady state and instead planetary climate experiences a limit cycle with four stages. Most of the time is spent with the planet in the Snowball state, where it warms very slowly as a result of CO$_2$ outgassing.}
\label{fig:cycles}
\end{figure}

\section{Climate cycles when the Snowball weathering is set to zero}
\label{sec:cycles}

\begin{figure}[ht!]
\begin{center}
\epsfig{file=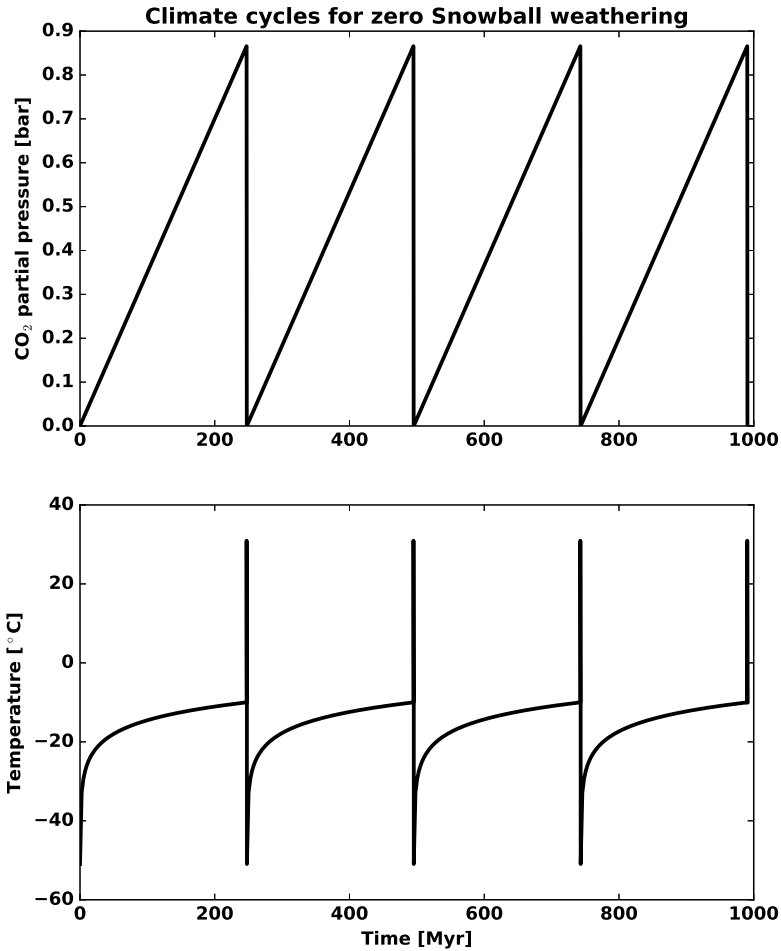, width=0.7\textwidth} 
\end{center}
\caption{Timeseries of CO$_2$ partial pressure and temperature for the
  limit cycle depicted in phase space in Figure~\ref{fig:cycles}. Most
  of the limit cycle is spent in the Snowball state, during which the
  CO$_2$ increases linearly in the atmosphere. I have made this plot
  based on the analytical expressions in section~\ref{sec:cycles}:
  exploiting the constant CO$_2$ accumulation in the Snowball state, a
  warm state CO$_2$ given by Equation~(\ref{eq:co2_bal2_soln}),
  temperature jumps at constant CO$_2$ for the transitions between
  states, and using Equations~(\ref{eq:warm_soln_vary}) and
  (\ref{eq:cold_soln_vary}) to find the temperatures from the CO$_2$
  values.}
\label{fig:cycles2}
\end{figure}

Alternatively, we can consider the situation where the weathering rate
is smaller than the CO$_2$ outgassing rate for temperatures less than
the temperature at which the planet transitions between the two
climate states ($T<T_i$). For simplicity, I will set the weathering
rate to zero for $T<T_i$, following \citet{Menou2015}. In this case no
Snowball steady state is possible because there is no weathering term
to balance CO$_2$ outgassing when the planet is experiencing a
Snowball (Equation~(\ref{eq:co2_bal})). Instead CO$_2$ simply
accumulates in the Snowball state, warming it, until the temperature
$T_i$ is reached. This causes the albedo to decrease (physically the
ice melts) and the planet abruptly jumps into the warm climate state.
If we assume that the CO$_2$ outgassing rate is low enough that no
warm climate steady state exists, then the warm climate leads to the
rapid removal of CO$_2$ by weathering until $T_i$ is again reached,
then the climate abruptly jumps into the Snowball state, and the cycle
repeats. Figure~\ref{fig:cycles} shows a diagram of this cycle in
phase space and Figure~\ref{fig:cycles2} shows timeseries of CO$_2$
partial pressure and temperature through the cycle.

The two transitions between the Snowball and warm states occur on the
timescale of relaxation back to the temperature tendency nullcline.
This timescale is given by $\frac{C}{a}\approx 3$ years, which is
essentially instantaneous for our purposes (the system is extremely
stiff). This allows us to make the approximation that as the CO$_2$
changes in either the warm or Snowball state, the climate exists along
the temperature tendency nullcline ($\frac{dT}{dt}$=0 in
Equation~(\ref{eq:climate_model})). The CO$_2$ partial pressure in the
warm state ($\tilde{P_w}$) as a function of the temperature in the
warm state ($\tilde{T_w}$) is therefore
\begin{equation}
  b\log\left(\frac{\tilde{P_w}}{P_0}\right)=a(\tilde{T_w}-T_0)+\frac{S_0}{4}(1-\alpha_w)-\frac{S}{4}(1-\alpha_w),
\label{eq:warm_soln_vary}
\end{equation}
and the CO$_2$ partial pressure in the cold state ($\tilde{P_c}$) as a
function of the temperature in the cold state ($\tilde{T_c}$) is
\begin{equation}
  b\log\left(\frac{\tilde{P_c}}{P_0}\right)=a(\tilde{T_c}-T_0)+\frac{S_0}{4}(1-\alpha_w)-\frac{S}{4}(1-\alpha_c).
\label{eq:cold_soln_vary}
\end{equation}
I have used a tilde for the CO$_2$ partial pressure and temperature
variables in these equations because they are not true solutions of
the system, since weathering never balances CO$_2$ outgassing during
the cycles.

$\frac{dP}{dt}$ is constant during the Snowball phase if the
weathering rate is zero (equal to $V$), so it is easy to calculate the time
spent in the Snowball phase ($\tau_c$) as
\begin{equation}
  \tau_c = \frac{\tilde{P_c}(T_i)-\tilde{P_w}(T_i)}{\frac{dP}{dt}} \approx \frac{\tilde{P_c}(T_i)}{\frac{dP}{dt}} = \frac{P_0}{V}e^{\frac{1}{b}(a(T_i-T_0)+\frac{S_0}{4}(1-\alpha_w)-\frac{S}{4}(1-\alpha_c))},
\label{eq:snow_time}
\end{equation}
where we have used the fact that in general
$\tilde{P_c}(T_i)\gg \tilde{P_w}(T_i)$. Similarly, if we assume that
silicate weathering is limited by the supply of silicate cations from
erosion \citep{mills11,foley2015role}, then we can approximate the
weathering rate as a constant during the warm phase. Using similar
logic as we used to get Equation~(\ref{eq:snow_time}), we arrive at a first
estimate for the time spent in the warm state ($\tau_{w1}$) of
\begin{equation}
  \tau_{w1} = \frac{\tilde{P_c}(T_i)-\tilde{P_w}(T_i)}{\frac{dP}{dt}} \approx \frac{\tilde{P_c}(T_i)}{\phi W_0-V} \approx \frac{\tilde{P_c}(T_i)}{\phi W_0} = \gamma \tau_c,
\label{eq:tau_w1}
\end{equation}
where $\phi$ is the factor by which the supply-limited maximum
weathering rate exceeds modern Earth's weathering rate
\citep[$\sim$2.5 is a good guess for an Earth-like planet,][]{mills11}
and $\gamma$ is the fractional reduction in warm state relative to
cold state lifetime due to the fact that the weathering is higher in
the warm state. Using $W_0$=20$V$, as in Figure~\ref{fig:cycles}, we
get $\gamma$=$\frac{1}{50}$. This would indicate that a negligible
amount of the time in the cycle is spent in the warm state relative to
the cold state.

It is more difficult to calculate the time spent in the warm climate
state if we let $\frac{dP}{dt}$ vary.
Substituting into Equation~(\ref{eq:co2_bal}) we find the following
differential equation for the CO$_2$ partial pressure in the warm
state ($\tilde{P_w}$)
\begin{equation}
\frac{d\tilde{P_w}}{dt}=V - W_0 e^{k(\tilde{T_w}-T_0)}\left( \frac{\tilde{P_w}}{P_0} \right)^\beta = V - W_0e^{\frac{k}{4a}(S-S_0)(1-\alpha_w)}\left( \frac{\tilde{P_w}}{P_0} \right)^{\beta+\frac{kb}{a}}.
\label{eq:co2_bal2}
\end{equation}
Equation~(\ref{eq:co2_bal2}) has a simple analytical solution for
$\beta+\frac{kb}{a}$=1, which happens to be the case for our default
parameters (Table~\ref{tab:params}). I will use this limit to illustrate the
behavior of the warm state CO$_2$ drawdown. The initial condition is
$\tilde{P_w}(t=0)=\tilde{P_c}(T_i)$, so that
Equation~(\ref{eq:co2_bal2}) is solved by
\begin{equation}
  \tilde{P_w}(t) = (\tilde{P_c}(T_i)-\frac{V}{W_0}P_0e^{\frac{k}{4a}(S_0-S)(1-\alpha_w)})e^{-\frac{W_0}{P_0}e^{\frac{k}{4a}(S-S_0)(1-\alpha_w)}t}+\frac{V}{W_0}P_0e^{\frac{k}{4a}(S_0-S)(1-\alpha_w)}.
\label{eq:co2_bal2_soln}
\end{equation}
We are seeking the time, $\tau_w$, such that
$\tilde{P_w}(t=\tau_w)=\tilde{P_w}(T_i)$, so plugging into
Equation~(\ref{eq:co2_bal2_soln}) we find
\begin{equation}
\tau_{w2} = \frac{P_0}{W_0}e^{\frac{k}{4a}(S_0-S)(1-\alpha_w)}\log \left( \frac{\tilde{P_c}(T_i)-\frac{V}{W_0}P_0e^{\frac{k}{4a}(S_0-S)(1-\alpha_w)}}{\tilde{P_w}(T_i)-\frac{V}{W_0}P_0e^{\frac{k}{4a}(S_0-S)(1-\alpha_w)}} \right).
\label{eq:tau_w2}
\end{equation}
In general $\tau_c \gg \tau_{w1} > \tau_{w2}$, as we would expect
because the temperatures are high and the weathering is fast in the
warm state (leading to a low $\tau_{w2}$) and we limit the weathering
rate in our other warm state timescale estimate ($\tau_{w1}$). For
example, for the parameters used in Figure~(\ref{fig:cycles}),
$\tau_c$=250~Myr (slow), the two estimates for the time spent in the
warm state are $\tau_{w1}$=5~Myr and $\tau_{w2}$=0.5~Myr (fast), and
the time to transition between the warm and Snowball states
($\frac{C}{a}$) is about 3 years (very fast).

Since the other components of the cycle take are short, $\tau_c$
(Equation~(\ref{eq:snow_time})) yields a good approximation of the
period of the total cycle, which we will call $\tau$. We can drop the
constants associated with the reference state in
Equation~(\ref{eq:snow_time}) as follows to think about the variable
dependencies of $\tau$
\begin{equation}
  \tau \propto  \frac{1}{V}e^{\frac{a}{b}T_i-\frac{S}{4b}(1-\alpha_c)}.
\label{eq:snow_time2}
\end{equation}
As one might expect, the period of the cycles is inversely
proportional to the CO$_2$ outgassing rate. The more interesting
aspect of Equation~(\ref{eq:snow_time2}) is that it gives us a
functional form for the dependence of the period of the cycle on the
stellar flux and albedo of the cold state. If we start in the warm
state and decrease the stellar flux (say by moving the planet away
from the star), a limit cycle will suddenly appear when the threshold
for existence of the warm state is crossed
(Equation~(\ref{eq:s_star})). If we continue to decrease the stellar
flux, the period of this cycle will grow exponentially as the stellar
flux is decreased. The exponential functional dependence ultimately
derives from the logarithmic effect of CO$_2$ on infrared emission to
space (Equation~(\ref{eq:climate_model})). This means that the period
of the cycle for planets near the outer edge of the habitable zone
will be very long, so the difference between having a permanent
Snowball state as in section~\ref{sec:stable_snow} and having very
slow cycles through the Snowball and warm states as described in this
section will be marginal. Similarly, the timescales will be
exponentially shorter as we decrease the Snowball albedo. Since ice
and snow have a much lower albedo for an M-star spectrum
\citep{Joshi:2012hu,shields2013effect}, this implies that the climate
cycles for planets in M-star systems would have much shorter periods.
Finally, increasing $b$ decreases the effect of changing the stellar
flux and Snowball albedo. It is important to note that since we have
set the weathering rate to zero in the Snowball, which is the part of
the cycle that determines its period, none of the  uncertain
weathering parameters affect the period of the cycle.

The dependence of $\tau$ on $T_i$ (Equation~(\ref{eq:snow_time2}) can
significantly affect the comparison of different models. For example,
I implemented a smoothed version of the albedo transition
(Equation~(\ref{eq:albedo})) and integrated the system numerically. I
found that the period of the cycle was strongly dependent on the
temperature smoothing of the albedo parameterization because this
affected the effective temperature at which the transition from the
Snowball to the warm climate occurred. Using $\frac{a}{b}=0.2$, a
decrease in $T_i$ by about 10~K leads to a decrease in $\tau$ by about
an order of magnitude. This sensitivity will affect the comparison of
cycle periods between models, however, the dependencies shown in
Equation~(\ref{eq:snow_time2}) should hold within a given model.

\begin{figure}[h!]
\begin{center}
\epsfig{file=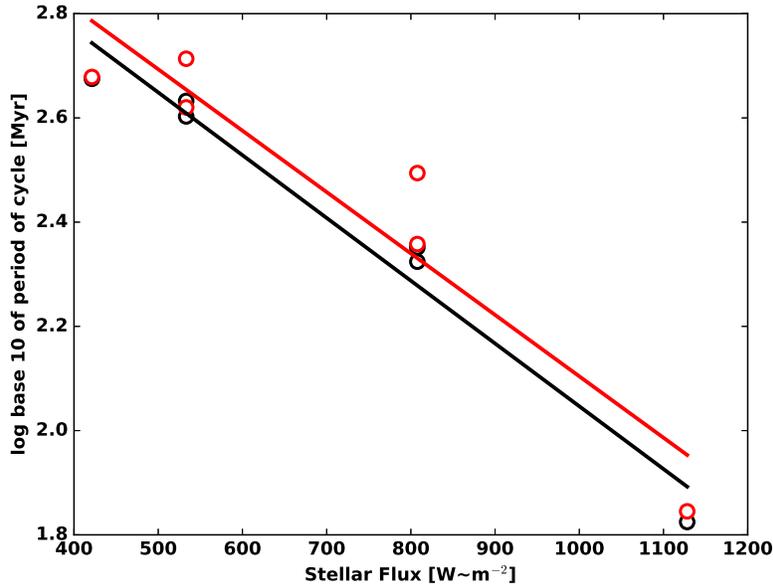, width=0.7\textwidth} 
\end{center}
\caption{Logarithm of the period of the climate cycles from
  \citet{Menou2015} (red circles) and time in the Snowball state
  (black circles) as a function of stellar flux. Lines of best fit,
  with corresponding colors, are also shown. Two of the simulations
  from \citet{Menou2015} have a CO$_2$ outgassing rate three times
  higher than the other simulations, and I adjusted the period of
  these simulations by multiplying it by three. }
\label{fig:menou}
\end{figure}

We can use Equation~(\ref{eq:snow_time2}) to understand simulations in
more complex models. For example, \citet{Menou2015} performed six
different simulations of climate cycles for planets at different
orbital distances (corresponding to different stellar fluxes), CO$_2$
outgassing rates, and values of $\beta$. Two of these simulations have
a CO$_2$ outgassing rate three times higher than the others, and,
according to Equation~(\ref{eq:snow_time2}), I have adjusted their
period by multiplying it by three. I have plotted the logarithm of the
adjusted period as a function of stellar flux in
Figure~\ref{fig:menou}. I have also plotted the time spent in the
Snowball state for each simulation, which we can calculate because
\citet{Menou2015} gives the percentage of the cycle spent in the warm
state. The first thing to note is that the logarithm of the period is
fairly linear in stellar flux, consistent with
Equation~(\ref{eq:snow_time2}). If we assume that the Snowball albedo
is 0.7, the slope of the line corresponds to a value of $b$ of about
27~W~m$^{-2}$, which is the right order of magnitude. Two of the
simulations have periods 10--20\% longer than expected. These
simulations have a lower value of $\beta$, and therefore spend a
higher fraction of their cycle in the warm state \citep{Menou2015}. We
should also note that even the time spent in the Snowball state does
not fall exactly on a straight line, and it is not a single-valued
function of stellar flux. The reasons for this are: (1)
\citet{Menou2015} uses a much more complex radiation model in which
infrared emission to space is not simply a linear function of the
logarithm of CO$_2$, (2) \citet{Menou2015} calculates the Snowball
albedo including the effect of the amount of CO$_2$, so it is not a
constant, and (3) the weathering parameterization that
\citet{Menou2015} uses allows some weathering for temperatures
slightly below the Snowball deglaciation temperature threshold, which
can delay the deglaciation and causes different results for different
values of $\beta$. Despite these differences,
Equation~(\ref{eq:snow_time2}) does an excellent job of describing the
qualitative behavior of period of the cycles from the simulations of
\citet{Menou2015}.

\section{Discussion}
\label{sec:discussion}

A major advance of this paper is to derive an explicit formula for the
effective outer edge of the habitable zone (Equation \ref{eq:s_star}).
Although the weathering parameters in this formula are uncertain, it
could be incorporated into future probabilistic estimates of
habitability of discovered exoplanets, with appropriate prior
distributions placed on weathering parameters. Once a planet is beyond
this limit, I have found that it will either experience a permanent
Snowball state or long cycles between a Snowball and warm climate,
depending on whether weathering goes completely to zero during the
Snowball or note. Either way the habitability of the planet would be
greatly reduced. Also, the fact that the Snowball episodes Earth has
experienced did end does not imply that the weathering was zero during
them \citep{Lehir08}, since the stellar flux was relatively high
during these episodes.

\citet{kopparapu2014habitable} found that the outer edge of the
traditional habitable zone has only a small dependence on planet size,
but the effective outer edge of the habitable zone could strongly
depend on planet size. A larger planet will tend to have a higher rate
of volcanism, and presumably CO$_2$ outgassing, yet it will also have
a larger overburden pressure for a given volatile inventory
\citep{Kite:2009p2923}. These competing effects will determine how
planetary size affects CO$_2$ outgassing rate, and consequently
susceptibility to loss of the warm climate state inside the
traditional habitable zone. Moreover, the CO$_2$ outgassing rate
should decrease strongly with time \citep{Kite:2009p2923}, which
indicates that planets near the outer edge of the traditional
habitable zone are more likely to actually be habitable in younger
systems.

The albedo and thermal phase curve of an Earth-like planet could be
interrogated to determine whether it was in a Snowball or warm climate
state \citep{cowan12-thermal}. This might be possible for a planet
near the outer edge of the habitable zone with the \textit{James Webb
  Space Telescope} \citep{yang2013,koll-2015}, and would certainly be
possible with a future mission such as the \textit{High Definition
  Space Telescope} \citep{dalcanton2015cosmic}. New geochronological
data \citep{Condon2016} suggest that the Sturtian and Marinoan
Snowball Earth episodes had a combined duration of about 80~Myr, which
is about 10\% of the time since they occurred, implying that a roughly
10\% Snowball duty cycle could be realistic for an Earth-like planet.
Planets near the outer edge of the traditional habitable zone that
have a stable warm state but are perturbed away from it and into a
Snowball could take longer to warm up via CO$_2$ outgassing and will
therefore spend a somewhat higher percentage of their time as a
Snowball. If these planets are anything like Earth, however, they
should still spend the vast majority of their time in the warm climate
state. Planets outside the effective outer edge of the habitable zone,
on the other hand, should spend most of their time as a Snowball.
Therefore a measurement of the fraction of Earth-like planets in a
warm state as a function of position in the habitable zone would tell
us whether CO$_2$-outgassing limitations on the habitable zone are
important. A large increase in the average number of Earth-like
planets in a Snowball state near the outer edge of the habitable zone
would indicate that CO$_2$-outgassing limitations are important on
average. In this way astronomical measurements could increase our
understanding of weathering. Similarly, continued study of river
catchments, paleoclimate, and laboratory weathering analogs should
help improve our understanding of weathering, and therefore the
effective outer edge of the habitable zone, which will inform the
astronomical search for habitable exoplanets.

In this paper I have not included the effect of CO$_2$ on planetary
albedo. This means that if the weathering is set to zero in a Snowball
state, then the CO$_2$ can always build up to high enough values to
cause deglaciation and lead to climate cycles if the warm state does
not exist (section \ref{sec:cycles}). For Snowball states that require
tens of bars of CO$_2$ to deglaciate, increased shortwave scattering
by CO$_2$ could prevent deglaciation from ever occurring. If this were
the case, the Snowball state could be stable even if the CO$_2$ cycle
is not balanced. If deglaciaton does occur at a very high CO$_2$
level, then the atmospheric albedo might be so high that the change in
surface albedo has a minimal effect on the planetary albedo
\citep{Wordsworth:2011p3221}, so the warming associated with
deglaciation is minimal. In this case CO$_2$ could still be drawn down
due to high CO$_2$ concentrations (Equation~(\ref{eq:co2_bal})),
leading to a climate cycle, but the planet would likely spend more
time in the warm state.

Finally, we should note that even if the effective outer edge of the
habitable zone occurs at a significantly higher stellar flux than the
traditional outer edge, there should still be plenty of habitats for
life in the universe. If the habitable zone were cut in half, the
proportion of Sun-like stars hosting Earth-like potentially habitable
planets would go from $\sim$5\% to $\sim$2.5\% \citep{Petigura:2013},
which might have been considered an optimistic estimate before the
\textit{Kepler} mission. Moreover, recent work on H$_2$-greenhouse
planets suggests that habitable planets can exist even outside of the
traditional habitable zone
\citep{Stevenson99,Pierrehumbert:2011p3366,Wordsworth2012-transient,abbot2015}.
Additionally, life, and maybe even animal life, seems to have survived
the Snowball Earth episodes, and could potentially survive permanent
or cyclical Snowball climates near the outer edge of the habitable
zone, although such conditions would certainly be less favorable to
complex life than modern Earth. Finally, if simple life can exist in
subglacial oceans on distant or unbound Earth-like planets
\citep{Laughlin:2000p2278,Abbot11}, then the Snowball planets beyond
the effective habitable zone would still be viable hosts for some sort
of life.

\section{Conclusions}
\label{sec:conclusions}

The main conclusions of this paper are:
\begin{enumerate}

\item The stellar flux at the effective outer edge of the habitable zone can be approximated by the following formula:
  \[S^\ast =
    S_0-\left(\frac{4}{1-\alpha_w}\right)\left(\frac{b}{\beta}\log\left(\frac{V}{W_0}\right)+k(T_0-T_i)(\frac{b}{\beta}+\frac{a}{k})\right).\]
  Larger values of $k$, the weathering-temperature rate constant,
  linearly decrease the stellar flux of the effective outer edge of
  the habitable zone, moving it farther from the star and providing
  more habitable space in the system. Smaller values of $\beta$, the
  weathering-CO$_2$ power law exponent, directly decrease the stellar
  flux of the effective outer edge of the habitable zone and also
  leverage the effect of increases in the CO$_2$ outgassing rate. If
  $k$ is increased by about a factor of two or $\beta$ is decreased by
  a factor of two, or some smaller combination of the two, then the
  effective outer edge of the habitable zone equals the traditional
  outer edge of the habitable zone even for modern Earth's CO$_2$
  outgassing rate, and none of the habitable zone would be lost. These
  changes are within the uncertainty in the values of $k$ and $\beta$.
  This equation also tells us that M-star planets should tend to have
  less of a reduction in their habitable zone due to limited CO$_2$
  outgassing, since $\alpha_w$, the warm state albedo, will tend to be
  smaller for M-star planets (making $S^\ast$ smaller).

  The formula for $S^\ast$ could be incorporated into probabilistic
  estimates of whether a discovered exoplanet is habitable, using
  appropriate priors on $k$ and $\beta$. It could similarly be used to
  estimate the fraction of stars that host an Earth-like planet given
  statistics of exoplanet occurrences. Both of these uses would aid in
  the planning of telescopes that would observe Earth-like planets and
  search for biosignatures.

\item Beyond the effective outer edge of the habitable zone (but
  inside the traditional outer edge) cycles between a Snowball and
  warm climate are possible if weathering is weak enough that the
  CO$_2$ needed to deglaciate a Snowball is reached before weathering
  can balance CO$_2$ outgassing (for example if the weathering rate is
  simply set to zero in a Snowball climate). If weathering occurs
  either subglacially or at the seafloor, it is possible to have a
  stable, attracting Snowball climate state.

\item If climate cycles between a Snowball and warm state occur, then
  the period of these cycles scales as
  \[ \tau \propto
    \frac{1}{V}e^{\frac{a}{b}T_i-\frac{S}{4b}(1-\alpha_c))}.\]
  This formula comes from the time spent in the Snowball state, which
  dominates the total period and can be calculated by dividing the
  CO$_2$ needed to deglaciate the Snowabll by the CO$_2$ outgassing
  rate (which explains why the period of the cycles is inversely
  proportional to the CO$_2$ outgassing rate). The exponential
  dependence on the temperature at which the Snowball state
  deglaciates and the negative exponential dependence on the stellar
  flux ultimately derive from the fact that CO$_2$ has a logarithmic
  effect on infrared emission to space and the greenhouse warming of a
  planet. The negative exponential dependence on stellar flux
  indicates that cycles near the outer edge of the habitable zone will
  have very long periods, and may be hard to distinguish from
  permanent Snowball states. The exponential dependence on the planetary
  albedo of the cold state indicates that climate cycles will have a
  shorter period for planets orbiting M-stars because the albedo of
  ice and snow is lower for an M-star spectrum. Finally, it is
  important to note that none of the uncertain weathering parameters
  appear in this scaling.

\end{enumerate}

\section{Acknowledgements}
I acknowledge support from the NASA Astrobiology Institute’s Virtual
Planetary Laboratory, which is supported by NASA under cooperative
agreement NNH05ZDA001C. I thank Navah Farahat for helping me learn
python plotting routines. I thank Cael Berry, Edwin Kite, Daniel Koll,
Mary Silber, and Robin Wordsworth for reading an early draft of this
paper and providing detailed and insightful suggestions.


\begin{thebibliography}{59}
\expandafter\ifx\csname natexlab\endcsname\relax\def\natexlab#1{#1}\fi

\bibitem[{Abbot(2015)}]{abbot2015}
Abbot, D.~S. 2015, Astrophysical Journal Letters, 815, {L3}

\bibitem[{Abbot {et~al.}(2012)Abbot, Cowan, \& Ciesla}]{abbot12-weathering}
Abbot, D.~S., Cowan, N.~B., \& Ciesla, F.~J. 2012, Astrophysical Journal, 756,
  178, {doi:10.1088/0004-637X/756/2/178}

\bibitem[{Abbot \& Switzer(2011)}]{Abbot11}
Abbot, D.~S., \& Switzer, E.~R. 2011, Astrophysical Journal, 735, L27,
  {doi:10.1088/2041-8205/735/2/L27}

\bibitem[{Abbot {et~al.}(2011)Abbot, Voigt, \&
  Koll}]{Abbot-et-al-2011:Jormungand}
Abbot, D.~S., Voigt, A., \& Koll, D. 2011, Journal of Geophysical Research,
  116, {D18103, doi:10.1029/2011JD015927}

\bibitem[{Berner(1994)}]{Berner:1994p3295}
Berner, R. 1994, Am J Sci, 294, 56

\bibitem[{Berner(2004)}]{Berner2004}
Berner, R.~A. 2004, The Phranerozoic Carbon Cycle (Oxford University Press, New
  York, N.Y.)

\bibitem[{Condon {et~al.}(2016)Condon, Macdonald, Rooney, Zhu, Schmitz, \&
  Bowring}]{Condon2016}
Condon, D., Macdonald, F.~A., Rooney, A.~D., Zhu, M., Schmitz, M.~D., \&
  Bowring, S.~A. 2016, Sci. Adv., submitted

\bibitem[{Cowan {et~al.}(2012)Cowan, Abbot, \& Voigt}]{cowan12-thermal}
Cowan, N.~B., Abbot, D.~S., \& Voigt, A. 2012, Astrophysical Journal, 757, {80,
  doi:10.1088/0004-637X/757/1/80}

\bibitem[{Dalcanton {et~al.}(2015)Dalcanton, Seager, Aigrain, Battel, Brandt,
  Conroy, Feinberg, Gezari, Guyon, Harris, {et~al.}}]{dalcanton2015cosmic}
Dalcanton, J., {et~al.} 2015, arXiv preprint arXiv:1507.04779

\bibitem[{Dasgupta(2013)}]{dasgupta2013ingassing}
Dasgupta, R. 2013, Rev Mineral Geochem, 75, 183

\bibitem[{Foley(2015)}]{foley2015role}
Foley, B.~J. 2015, The Astrophysical Journal, 812, 36

\bibitem[{Goldblatt {et~al.}(2013)Goldblatt, Robinson, Zahnle, \&
  Crisp}]{Goldblatt:2013}
Goldblatt, C., Robinson, T.~D., Zahnle, K.~J., \& Crisp, D. 2013, Nature
  Geoscience, 6, 661

\bibitem[{Goldblatt \& Watson(2012)}]{Goldblatt:2012}
Goldblatt, C., \& Watson, A.~J. 2012, Philosophical Transactions Of The Royal
  Society A-Mathematical Physical And Engineering Sciences, 370, 4197

\bibitem[{Gough(1981)}]{GOUGH:1981p2371}
Gough, D.~O. 1981, Sol Phys, 74, 21

\bibitem[{Grott {et~al.}(2011)Grott, Morschhauser, Breuer, \&
  Hauber}]{grott2011volcanic}
Grott, M., Morschhauser, A., Breuer, D., \& Hauber, E. 2011, Earth and
  Planetary Science Letters, 308, 391

\bibitem[{Halevy \& Head~III(2014)}]{halevy2014episodic}
Halevy, I., \& Head~III, J.~W. 2014, Nature Geoscience, 7, 865

\bibitem[{Haqq-Misra {et~al.}(2016)Haqq-Misra, Kopparapu, Batalha, Harman, \&
  Kasting}]{haqq2016limit}
Haqq-Misra, J., Kopparapu, R.~K., Batalha, N.~E., Harman, C.~E., \& Kasting,
  J.~F. 2016, arXiv preprint arXiv:1605.07130

\bibitem[{Hoffman {et~al.}(1998)Hoffman, Kaufman, Halverson, \&
  Schrag}]{Hoffman98}
Hoffman, P.~F., Kaufman, A.~J., Halverson, G.~P., \& Schrag, D.~P. 1998,
  Science, 281, 1342

\bibitem[{Huybers \& Langmuir(2009)}]{huybers2009feedback}
Huybers, P., \& Langmuir, C. 2009, Earth and Planetary Science Letters, 286,
  479

\bibitem[{Joshi \& Haberle(2012)}]{Joshi:2012hu}
Joshi, M.~M., \& Haberle, R.~M. 2012, Astrobiology, 12, 3

\bibitem[{Kadoya \& Tajika(2014)}]{Kadoya:2014kd}
Kadoya, S., \& Tajika, E. 2014, The Astrophysical Journal, 790, 107

\bibitem[{Kasting(1988)}]{Kasting88}
Kasting, J.~F. 1988, Icarus, 74, 472

\bibitem[{Kasting {et~al.}(1993)Kasting, Whitmire, \& Reynolds}]{Kasting93}
Kasting, J.~F., Whitmire, D.~P., \& Reynolds, R.~T. 1993, Icarus, 101, 108

\bibitem[{Kirschvink(1992)}]{Kirschvink92}
Kirschvink, J. 1992, in {The Proterozoic Biosphere: A Multidisciplinary Study},
  ed. J.~Schopf \& C.~Klein (Cambridge University Press, New York), 51--52

\bibitem[{Kite {et~al.}(2009)Kite, Manga, \& Gaidos}]{Kite:2009p2923}
Kite, E., Manga, M., \& Gaidos, E. 2009, Astrophys Journal, 700, 1732,
  {doi:10.1088/0004-637X/700/2/1732}

\bibitem[{Kite {et~al.}(2015)Kite, Howard, Lucas, Armstrong, Aharonson, \&
  Lamb}]{kite2015stratigraphy}
Kite, E.~S., Howard, A.~D., Lucas, A.~S., Armstrong, J.~C., Aharonson, O., \&
  Lamb, M.~P. 2015, Icarus, 253, 223

\bibitem[{Koll \& Abbot(2015)}]{koll-2015}
Koll, D.~D.~B., \& Abbot, D.~S. 2015, The Astrophysical Journal, 802, 21,
  {doi:10.1088/0004-637X/802/1/21}

\bibitem[{Kopparapu(2013)}]{Kopparapu:2013fp}
Kopparapu, R.~K. 2013, Astrophysical Journal Letters, 767, L8

\bibitem[{Kopparapu {et~al.}(2014)Kopparapu, Ramirez, SchottelKotte, Kasting,
  Domagal-Goldman, \& Eymet}]{kopparapu2014habitable}
Kopparapu, R.~K., Ramirez, R.~M., SchottelKotte, J., Kasting, J.~F.,
  Domagal-Goldman, S., \& Eymet, V. 2014, The Astrophysical Journal Letters,
  787, L29

\bibitem[{Kopparapu {et~al.}(2013)Kopparapu, Ramirez, Kasting, Eymet, Robinson,
  Mahadevan, Terrien, Domagal-Goldman, Meadows, \& Deshpande}]{Kopparapu:2013}
Kopparapu, R.~K., {et~al.} 2013, The Astrophysical Journal, 765, 131

\bibitem[{Laughlin \& Adams(2000)}]{Laughlin:2000p2278}
Laughlin, G., \& Adams, F. 2000, Icarus, 145, 614

\bibitem[{Le~{H}ir {et~al.}(2008)Le~{H}ir, Ramstein, Donnadieu, \&
  Godderis}]{Lehir08}
Le~{H}ir, G., Ramstein, G., Donnadieu, Y., \& Godderis, Y. 2008, Geology, 36,
  47

\bibitem[{Leconte {et~al.}(2013{\natexlab{a}})Leconte, Forget, Charnay,
  Wordsworth, \& Pottier}]{leconte2013increased}
Leconte, J., Forget, F., Charnay, B., Wordsworth, R., \& Pottier, A.
  2013{\natexlab{a}}, Nature, 504, 268

\bibitem[{Leconte {et~al.}(2013{\natexlab{b}})Leconte, Forget, Charnay,
  Wordsworth, Selsis, Millour, \& Spiga}]{Leconte:2013gv}
Leconte, J., Forget, F., Charnay, B., Wordsworth, R., Selsis, F., Millour, E.,
  \& Spiga, A. 2013{\natexlab{b}}, Astronomy And Astrophysics, 554, A69

\bibitem[{Menou(2015)}]{Menou2015}
Menou, K. 2015, Earth and Planetary Science Letters, 429, 20

\bibitem[{Mills {et~al.}(2011)Mills, Watson, Goldblatt, Boyle, \&
  Lenton}]{mills11}
Mills, B., Watson, A.~J., Goldblatt, C., Boyle, R., \& Lenton, T.~M. 2011,
  Nature Geosciences, 4, 861, {DOI: 10.1038/NGEO1305}

\bibitem[{Nakajima {et~al.}(1992)Nakajima, Hayashi, \& Abe}]{Nakajima92}
Nakajima, S., Hayashi, Y.~Y., \& Abe, Y. 1992, Journal of the Atmospheric
  Sciences, 49, 2256

\bibitem[{Petigura {et~al.}(2013)Petigura, Howard, \& Marcy}]{Petigura:2013}
Petigura, E.~A., Howard, A.~W., \& Marcy, G.~W. 2013, Proceedings of the
  National Academy of Sciences, 110, 19273

\bibitem[{Pierrehumbert \& Gaidos(2011)}]{Pierrehumbert:2011p3366}
Pierrehumbert, R., \& Gaidos, E. 2011, Astrophys J Lett, 734, L13,
  {doi:10.1088/2041-8205/734/1/L13}

\bibitem[{Pierrehumbert(2010)}]{Pierrehumbert:2010-book}
Pierrehumbert, R.~T. 2010, Principles of Planetary Climate (Cambridge
  University Press)

\bibitem[{Rose(2015)}]{rose2015stable}
Rose, B.~E. 2015, Journal of Geophysical Research: Atmospheres, 120, 1404

\bibitem[{Shields {et~al.}(2013)Shields, Meadows, Bitz, Pierrehumbert, Joshi,
  \& Robinson}]{shields2013effect}
Shields, A.~L., Meadows, V.~S., Bitz, C.~M., Pierrehumbert, R.~T., Joshi,
  M.~M., \& Robinson, T.~D. 2013, Astrobiology, 13, 715

\bibitem[{{Stevenson}(1999)}]{Stevenson99}
{Stevenson}, D.~J. 1999, Nature, 400, 32, {doi:10.1038/21811}

\bibitem[{Strogatz(1994)}]{Strogatz-1994:nonlinear}
Strogatz, S. 1994, Nonlinear dynamics and chaos (Westview Press)

\bibitem[{Tajika(2007)}]{Tajika2007}
Tajika, E. 2007, Earth, 59, 293

\bibitem[{Voigt \& Abbot(2012)}]{voigt12-dynamics}
Voigt, A., \& Abbot, D.~S. 2012, Climate of the Past, 8, 2079,
  {doi:10.5194/cp-8-2079-2012}

\bibitem[{Voigt {et~al.}(2011)Voigt, Abbot, Pierrehumbert, \&
  Marotzke}]{voigt11}
Voigt, A., Abbot, D.~S., Pierrehumbert, R.~T., \& Marotzke, J. 2011, Clim.
  Past, 7, 249, doi:10.5194/cp-7-249-2011

\bibitem[{Walker {et~al.}(1981)Walker, Hays, \&
  Kasting}]{Walker-Hays-Kasting-1981:negative}
Walker, J.~C.~G., Hays, P.~B., \& Kasting, J.~F. 1981, Journal of Geophysical
  Research, 86, 9776

\bibitem[{West {et~al.}(2005)West, Galy, \& Bickle}]{west2005tectonic}
West, A.~J., Galy, A., \& Bickle, M. 2005, Earth and Planetary Science Letters,
  235, 211

\bibitem[{Wolf \& Toon(2014)}]{Wolf:2014}
Wolf, E., \& Toon, O. 2014, Geophysical Research Letters, 41, 167–172, dOI:
  10.1002/2013GL058376

\bibitem[{Wolf \& Toon(2015)}]{wolf2015evolution}
---. 2015, Journal of Geophysical Research, 120, 5775

\bibitem[{Wordsworth(2012)}]{Wordsworth2012-transient}
Wordsworth, R. 2012, Icarus, 219, 267, {doi:10.1016/j.icarus.2012.02.035}

\bibitem[{Wordsworth {et~al.}(2013)Wordsworth, Forget, Millour, Head,
  Madeleine, \& Charnay}]{Wordsworth:2013fk}
Wordsworth, R., Forget, F., Millour, E., Head, J.~W., Madeleine, J.~B., \&
  Charnay, B. 2013, Icarus, 222, 1

\bibitem[{Wordsworth {et~al.}(2011)Wordsworth, Forget, Selsis, Millour,
  Charnay, \& Madeleine}]{Wordsworth:2011p3221}
Wordsworth, R.~D., Forget, F., Selsis, F., Millour, E., Charnay, B., \&
  Madeleine, J.-B. 2011, Astrophys J Lett, 733, L48,
  {doi:10.1088/2041-8205/733/2/L48}

\bibitem[{Yang {et~al.}(2014)Yang, Bou\'e, Fabrycky, \& Abbot}]{yang2014}
Yang, J., Bou\'e, G., Fabrycky, D.~C., \& Abbot, D.~S. 2014, Astrophysical
  Journal Letters, 787, {L2, doi:10.1088/2041-8205/787/1/L2}

\bibitem[{Yang {et~al.}(2013)Yang, Cowan, \& Abbot}]{yang2013}
Yang, J., Cowan, N.~B., \& Abbot, D.~S. 2013, Astrophysical Journal Letters,
  771, {L45, DOI:10.1088/2041-8205/771/2/L45}

\bibitem[{Yang {et~al.}(2012{\natexlab{a}})Yang, Peltier, \& Hu}]{yang2011a}
Yang, J., Peltier, W., \& Hu, Y. 2012{\natexlab{a}}, Journal of Climate, 25,
  2711, {doi: 10.1175/JCLI-D-11-00189.1}

\bibitem[{Yang {et~al.}(2012{\natexlab{b}})Yang, Peltier, \& Hu}]{yang2011b}
---. 2012{\natexlab{b}}, Journal of Climate, 25, 2737, {doi:
  10.1175/JCLI-D-11-00190.1}

\bibitem[{Yang {et~al.}(2012{\natexlab{c}})Yang, Peltier, \& Hu}]{yang2012}
Yang, J., Peltier, W.~R., \& Hu, Y. 2012{\natexlab{c}}, Climate of the Past, 8,
  907, {doi:10.5194/cp-8-907-2012}

\end{thebibliography}
\end{document}